\documentclass[lettersize,journal]{IEEEtran}

\usepackage{amsmath,amsfonts}
\usepackage{array}
\usepackage{textcomp}
\usepackage{stfloats}
\usepackage{url}
\usepackage{verbatim}
\usepackage{graphicx}
\usepackage{svg}
\usepackage{float}
\usepackage{tabularx}

\usepackage{booktabs}
\usepackage{multirow}

\usepackage[caption=false,font=normalsize,labelfont=sf,textfont=sf]{subfig}

\usepackage{enumitem}

\usepackage{algorithm}
\usepackage{algorithmic}

\def\BibTeX{{\rm B\kern-.05em{\sc i\kern-.025em b}\kern-.08em
    T\kern-.1667em\lower.7ex\hbox{E}\kern-.125emX}}
\usepackage{balance}

\begin{document}

\title{A Positron Range Correction with Texture Preservation Framework in PET Imaging}
\author{N.~Encina-Baranda,~\IEEEmembership{Member,~IEEE,}
        Y.~Zheng,~\IEEEmembership{Member,~IEEE,}
        J.~Cabello,~\IEEEmembership{Member,~IEEE,}
        R.~J.~Paneque-Yunta,~\IEEEmembership{Member,~IEEE,}
        C.~M.~Solano-Cordero,
        A.~Lopez-Montes,~\IEEEmembership{Member,~IEEE,}
        M.~Conti,~\IEEEmembership{Member,~IEEE,}
        J.~L.~Herraiz,~\IEEEmembership{Member,~IEEE}%
\thanks{This work involved human subjects in its research. All patient data used in this study were obtained from Washington University, St. Louis, MO, USA. The study was approved by the Washington University Institutional Review Board (IRB) and was performed in line with the Declaration of Helsinki. All patients provided written informed consent.
}
\thanks{N. Encina-Baranda, R. J. Paneque-Yunta, C.M Solano-Cordero and J. L. Herraiz are with the Nuclear Physics Group and IPARCOS, Universidad Complutense de Madrid, Madrid, Spain, and also with the Instituto de Investigacion Sanitaria, Hospital Clinico San Carlos (IdISSC), Madrid, Spain (e-mail: nencina@ucm.es; rpaneque@ucm.es; cindysol@ucm.es; jlopezhe@ucm.es).}
\thanks{A. Lopez-Montes is with the Department of Nuclear Medicine, Bern University Hospital, Bern, Switzerland (e-mail: alejandro.lopez@unibe.ch).}%
\thanks{Y. Zheng, J. Cabello, and M. Conti are with Siemens Medical Solutions USA, Inc., Knoxville, TN, USA (e-mail: yifan.zheng@siemens-healthineers.com; jorge.cabello@siemens-healthineers.com; maurizioconti@siemens-healthineers.com).}}
\markboth{SUBMITTED TO IEEE TRPMS - UNDER REVIEW}%
{}

\maketitle

\begin{abstract}
Positron range (PR) blurring is a fundamental resolution limitation in PET imaging with high-energy positron emitters such as $^{82}$Rb, causing contrast loss and spill-out effects across heterogeneous tissue interfaces. We propose PRC-TP, a positron range correction (PRC) framework with explicit texture preservation that decouples deterministic resolution recovery from stochastic texture restoration. A nnFormer-based neural network (NN) was trained on patient-derived Monte Carlo simulations to map PR-degraded $^{82}$Rb reconstructions to PR-free references using attenuation maps as anatomical context. However, this NN also significantly removed the noise in the images, which could impact some texture analysis methods or make the images look unrealistic. An auxiliary Noise2Noise model estimates that smoothing effect, enabling texture extraction and transfer to the PR-corrected prediction through Model-consistent Texture Re-Injection (MTRI). In simulated patients, PRC-TP preserved contrast recovery close to ground truth (GT) (98.96--99.04\%) while restoring noise and CNR closer to the reference. The function-based MTRI formulation achieved near unity global texture amplitude agreement with GT ($0.997 \pm 0.011$), reducing the input texture amplitude bias ($0.951 \pm 0.011$). Radiomics analysis showed improved agreement with GT across texture-sensitive feature families. A clinical $^{82}$Rb evaluation showed trends consistent with simulations, including comparable contrast-ratio increase ($10.18\%$ vs. $10.99\%$) and restoration of texture suppressed by PRC. These results support PRC-TP as a practical framework for resolution recovery with acquisition-consistent texture preservation in PET imaging.
\end{abstract}


\section{Introduction}
\IEEEPARstart{P}{ositron} emission tomography (PET) is a quantitative
molecular imaging technique that provides three-dimensional images of
radiotracer distribution with high sensitivity \cite{vandenberghe2020state}
and has become indispensable in clinical oncology \cite{duclos2021pet},
cardiology \cite{pan2016clinical}, and neurology \cite{lu2015pet}. Despite
its widespread clinical adoption, PET image quality remains intrinsically
limited by several physical factors, including positron range (PR), which is
defined as the distance between the positron emission and its annihilation point
\cite{levin1999calculation}. It depends on the positron energy spectrum and
the stopping power of the surrounding medium \cite{paneque2025analytical},
and introduces a radionuclide and tissue dependent resolution degradation
that is independent of PET scanner hardware and reconstruction parameters
\cite{kemerink2011effect}. As shown in Table~\ref{tab:radionuclides},
higher-energy positron emitters exhibit longer ranges, particularly in
low-density media such as lung tissue.

\begin{table}[!h]
\centering
\caption{Physical properties, mean positron energies, mean positron ranges
($R_{\text{mean}}$), and maximum positron ranges ($R_{\max}$) of selected
radionuclides in water and lung tissue. The range values were obtained from
Monte Carlo simulations using the PeneloPET toolkit, as reported in
\cite{cal2013positron}.}
\label{tab:radionuclides}

\footnotesize
\setlength{\tabcolsep}{2.5pt}
\renewcommand{\arraystretch}{1.1}

\begin{tabular}{c c p{2.8cm} c c c}
\toprule
\textbf{Radionuclide} 
& \textbf{$T_{1/2}$} 
& \textbf{Mean $E_{\beta^+}$} 
& \textbf{Material} 
& \textbf{$R_{\text{mean}}$}
& \textbf{$R_{\max}$} \\

 & \textbf{(min)} 
 & \textbf{(keV)} 
 &  
 & \textbf{(mm)}
 & \textbf{(mm)} \\
\midrule

$^{18}$F  
& 109.8   
& 249.8 (97\%)             
& Water 
& 0.57 
& 2.16 \\
 &       
 &                          
 & Lung  
 & 1.85 
 & 7.49 \\

$^{13}$N  
& 10.0  
& 491.8 (99.8\%)           
& Water 
& 1.40 
& 4.88 \\
 &       
 &                          
 & Lung  
 & 4.61 
 & 16.1 \\

$^{68}$Ga  
& 67.7  
& 352.6 (1.2\%),   836 (87.9\%)           
& Water 
& 2.69 
& 9.06 \\
 &       
 &                          
 & Lung  
 & 8.86 
 & 27.1\\

$^{82}$Rb 
& 1.3   
& 1168 (13\%), 1535 (82\%) 
& Water 
& 5.33 
& 16.5 \\
 &       
 &                          
 & Lung  
 & 17.6 
 & 52.0 \\

\bottomrule
\end{tabular}
\end{table}

Modern digital clinical PET systems such as the Biograph Vision PET/CT
(Siemens Healthineers) achieve a transaxial spatial resolution of
approximately 3.5--3.6~mm FWHM at 1~cm radial offset from the center of the
field of view under the NEMA NU~2-2012 standard
\cite{van2019performance}. Consequently, for high-energy emitters such as
$^{68}$Ga and $^{82}$Rb, PR blurring can become comparable to the intrinsic
scanner resolution in water and may substantially exceed it in low-density
media. This effect can compromise spatial resolution and contrast recovery,
particularly across tissue interfaces.

A wide variety of positron range correction (PRC) strategies have been proposed, including both reconstruction-based and post-reconstruction approaches. Early reconstruction-based methods integrated PR modeling directly into the iterative reconstruction, including projection-domain deconvolution \cite{derenzo2007mathematical}, tissue-dependent PRC based on CT-derived segmentation \cite{cal2011study}, and subsequent extensions accounting for magnetic field effects in PET/MR \cite{kraus2012simulation}. More advanced tissue-dependent spatially variant PRC (TDSV-PRC) approaches employed anisotropic Monte Carlo–derived kernels scaled by local tissue properties, enabling the first clinical implementations in cardiac $^{82}$Rb PET \cite{cal2015tissue,kertesz2022implementation,kertesz2022positron,lassen2025positron}. More recently, hybrid physics--deep learning reconstruction methods such as dual-input dynamic convolution (DDConv) have been proposed, in which neural networks (NNs) predict voxel-wise anisotropic kernels from PET and CT information during forward and backprojection \cite{mellak2025dual}. While physically well grounded, these reconstruction-based methods remain computationally demanding and sensitive to noise amplification and ringing artifacts, limiting their robustness and clinical adoption \cite{cal2018improving,gavriilidis2022positron}.

To improve clinical practicality, post-reconstruction PRC methods have been also explored, including classical deconvolution techniques~\cite{rukiah2018investigation} and supervised deep learning approaches that map PR-degraded images to PR-corrected references, often incorporating attenuation maps~\cite{herraiz2020deep,encina2026tissue}. More broadly, recent transformer-based PET restoration methods have addressed related tasks such as simultaneous partial volume correction and denoising, showing the potential of deep learning (DL) to recover degraded PET images from low count or resolution-limited acquisitions~\cite{Kaviani2026}. Although these approaches improve contrast and resolution, they generally synthesize denoised images. 
As a result, these methods primarily focus on correcting deterministic PR effects, but do not explicitly decouple PRC from the smoothing of the stochastic texture. Self-supervised and generative strategies have been proposed to address texture and noise restoration~\cite{xie2025noise}; however, the restored texture in such approaches is typically inferred from the training data or model assumptions, rather than being explicitly extracted from the measured acquisition itself.

Under supervised training, NN outputs tend to exhibit smoothing, whereby stochastic fluctuations are suppressed (i.e. noise reduction) \cite{lehtinen2018noise2noise}. This mainly comes from three complementary factors: (i) the regression nature of the loss function (e.g., conditional mean for $L_2$ or median for $L_1$ loss), which promotes estimation of a central tendency; (ii) a data-driven effect due to the mismatch in noise realizations between input and target, which suppresses inconsistent high-frequency fluctuations; and (iii) a model-driven effect related to the intrinsic spectral bias of NNs toward low-frequency components \cite{rahaman2019spectral}. Together, these factors result in outputs with reduced high-frequency texture compared to real PET images. While generative models attempt to restore this variability, they may introduce hallucinated structures not supported by the underlying measurements, raising concerns about quantitative reliability \cite{cohen2018distribution, antun2020instabilities}.

In this work, we address this limitation by explicitly decoupling structural
resolution recovery from stochastic texture preservation in PET. Although the
method is demonstrated in $^{82}$Rb cardiac PET, the proposed formulation is not  restricted to a specific radionuclide. 

The proposed framework Positron Range Correction with Texture Preservation (PRC-TP) implements this decoupling through three complementary steps: (i) a tissue-guided DL model trained on patient-derived Monte Carlo (MC) simulations corrects PR blurring using PR-free images as reference, with anatomical guidance incorporated through attenuation maps and a tissue-dependent loss function; (ii) an auxiliary model is trained to reproduce the implicit smoothing behavior of the PRC network, yielding a smoothed representation consistent with the PRC output; and (iii) a deterministic step transfers high-frequency texture from the original PET image to the PR-corrected output while avoiding reintroduction of PR effects.

\vspace{-0.2cm}
\section{Proposed Framework}
To explicitly separate structural signal recovery from texture preservation, reconstructed PET images are approximated by the following model:
\begin{equation}
S(v) = \tilde{S}(v)\bigl(1 + \eta(v)\bigr),
\label{eq:eq_multiplicative}
\end{equation}
where $S(v)$ denotes a generic reconstructed PET image at voxel $v$, $\tilde{S}(v)$ represents the noise-free image, and $\eta(v)$ denotes zero-mean stochastic fluctuations arising from photon counting statistics and reconstruction-induced correlations \cite{barrett2013foundations}. This relative multiplicative formulation does not impose a specific reconstructed-image noise distribution, but provides a practical way to express PET texture.
\begin{table}[h]
\centering
\caption{Nomenclature. Summary of the symbols used to describe the PRC-TP framework.}
\label{tab:nomenclature}
\renewcommand{\arraystretch}{1.25}
\begin{tabularx}{\columnwidth}{@{}l X@{}}
\toprule
\textbf{Symbol} & \textbf{Description} \\
\midrule
\multicolumn{2}{@{}l}{\textbf{Images, maps, and domain}}\\
$v$ & Voxel index. \\
$\Omega$ & Image (spatial) domain. \\
$S(v)$ & Reconstructed PET image. \\
$\tilde{S}(v)$ & Noise-free signal component of $S$. \\
$\eta(v)$ & Zero-mean stochastic fluctuation. \\
$X$ & PR-degraded PET image (input). \\
$X'$ & Independent noisy realization of $X$. \\
$Y$ & PR-free ground truth (GT). \\
$\hat{Y}$ & PR-corrected image (PRC output). \\
$\tilde{X}$ & Smoothed estimate of $X$ (N2N output). \\
$\mu$ & Attenuation map. \\
$Y_{\mathrm{PRC\text{-}TP}}$ & Final texture-preserving image. \\
\addlinespace
\multicolumn{2}{@{}l}{\textbf{Operators and models}}\\
$\mathcal{F}_{\mathrm{PRC}}$ & PRC network operator. \\
$\mathcal{F}_{\mathrm{N2N}}$ & Noise2Noise smoothing operator. \\
\addlinespace
\multicolumn{2}{@{}l}{\textbf{Texture model}}\\
$\eta_X(v)$ & Relative texture from $X$. \\
$\eta_Y(v)$ & Relative texture from $Y$. \\
$\eta_{\mathrm{MTRI}}(v)$ & Re-injected texture. \\
$s$ & Scalar signal level. \\
$a,\,b,\,c,\,s_0$ & Texture-amplitude function parameters. \\
$\sigma_X(s)$ & Signal-dependent texture amplitude. \\
$A(v)$ & Local amplitude transfer factor. \\
$\Delta_{\mathrm{PR}}(v)$ & PR-change proxy. \\
\bottomrule
\end{tabularx}
\end{table}

Let \(X \in \mathcal{X}\) denote the measured PR-degraded PET image, \(Y \in \mathcal{Y}\) the corresponding PR-free ground truth (GT), \(\mu \in \mathcal{M}\) the attenuation map, and \(X'\) an independent noisy realization of \(X\) used for Noise2Noise training.

\begin{figure*}
    \centering
    \includegraphics[width=1\linewidth]{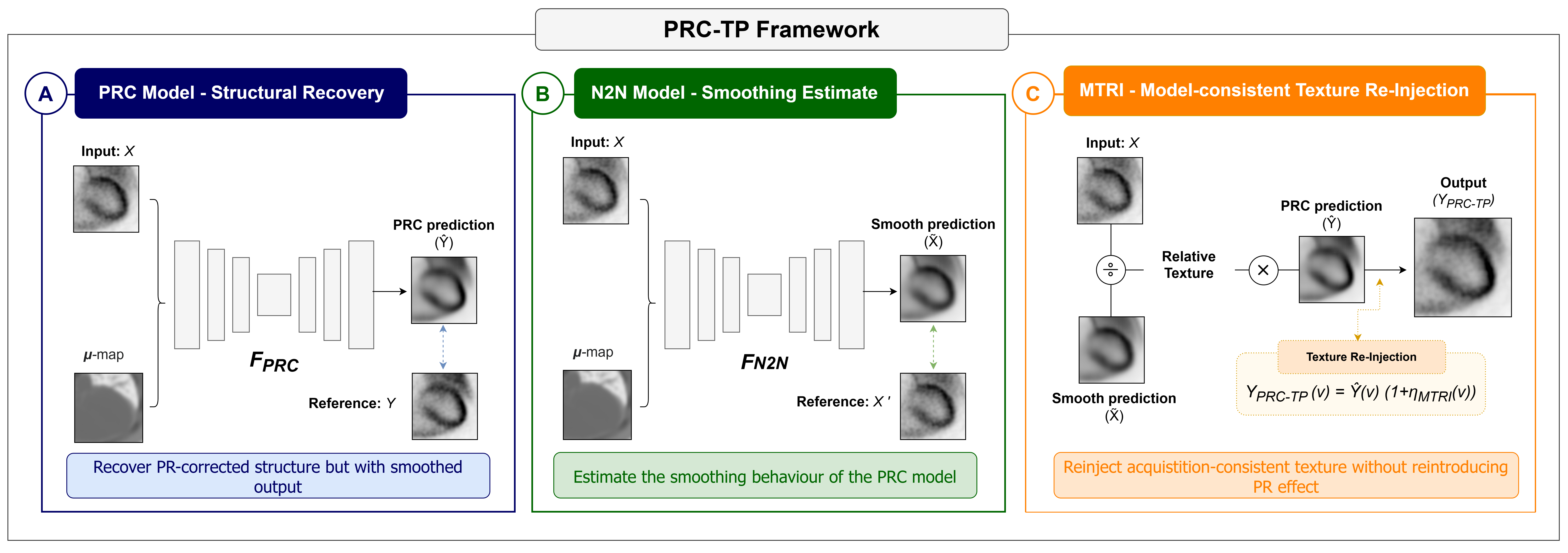}
    \caption{PRC-TP framework. \textbf{(A)} A deep learning PRC model estimates PR-corrected structure from PR-degraded PET and attenuation maps, producing a smoothed output. \textbf{(B)} A Noise2Noise (N2N) model reproduces the smoothing behavior of the PRC network using independent noisy realizations. \textbf{(C)} Model-consistent Texture Re-Injection (MTRI) restores acquisition-consistent texture via relative modulation.}
    \label{fig:framework}
\end{figure*}

Based on this notation, the proposed PRC-TP framework is decomposed into three sequential steps: PRC, smoothing estimation, and texture re-injection. A schematic overview of the method is shown in Fig.~\ref{fig:framework}.

\vspace{-0.15cm}
\subsection{Positron Range Correction (PRC)}
This step is formulated as
\begin{equation}
\hat{Y} = \mathcal{F}_{\text{PRC}}(X, \mu),
\label{eq:prc_mapping}
\end{equation}
where $\mathcal{F}_{\text{PRC}}$ is a NN–parameterized operator, and $\hat{Y}$ is the estimated PR-corrected image. In practice, the network is trained to learn the mapping 
\begin{equation}
    \bigl(X, \mu\bigl)  \;\longrightarrow\; Y.
\end{equation}
This training strategy ensures that input and target share the same underlying tracer distribution while differing in the presence of PR effects. The attenuation map $\mu$ is included to provide tissue information, as PR is material-dependent. 

As it was previously mentioned, due to the regression nature of the problem, the mismatch in noise realizations between $X$ and $Y$, together with the spectral bias of the network and the loss formulation toward central tendency, the learned mapping produces a smoother output $\hat{Y}$.

\vspace{-0.15cm}
\subsection{Smoothing Estimation}
This step is designed to reproduce the implicit smoothing behavior produced by $\mathcal{F}_{\text{PRC}}$. The Noise2Noise (N2N) model is defined as a smoothing operator, such that
\begin{equation}
\tilde{X} = \mathcal{F}_{\text{N2N}}\bigl(X, \mu\bigr),
\label{eq:n2n_mapping}
\end{equation}
where $X$ and $\tilde{X}$ denote the input image and its smoothed estimate. In contrast to the first step, the network is trained to learn the mapping 
\begin{equation}
\bigl(X, \mu\bigl) \;\longrightarrow\; X'.
\end{equation}
This formulation exploits the independence of noise realizations to preserve the shared structural content while attenuating stochastic fluctuations, yielding a smoother estimate that serves as a proxy for the underlying signal component $\tilde{S}(v)$ in Eq.~\ref{eq:eq_multiplicative}. For validation purposes, the same operator can also be applied to the GT image, yielding a smoothed version.

While a general Noise2Noise framework relies on the independence of noise realizations, the $\mathcal{F}_{\text{N2N}}$ operator is designed to reproduce the same smoothing behavior as the $\mathcal{F}_{\text{PRC}}$ by using the same architecture, sampling strategy, loss function, and input representation. The inclusion of the attenuation map $\mu$ ensures consistency of the learned smoothing behavior across both steps. Therefore, the only difference between both training schemes lies in the target image.
\vspace{-0.15cm}

\subsection{Model-consistent Texture Re-Injection (MTRI)}
Based on the model in Eq.~(\ref{eq:eq_multiplicative}) and the output obtained from $\mathcal{F}_{\mathrm{N2N}}$, we first define the relative texture of the input PET image $X(v)$ as
\begin{equation}
\eta_X(v) \approx \frac{X(v)}{\tilde{X}(v)} -1.
\label{eq:mti_modulation}
\end{equation}
This term captures the acquisition-specific stochastic fluctuation of the PET image relative to the local signal estimate. The same definition is applied to the GT to obtain its relative texture, $\eta_Y(v) \approx Y(v)/\tilde{Y}(v) - 1$, where $\tilde{Y} = \mathcal{F}_{\mathrm{N2N}}(Y,\mu)$ is the smoothed estimate of $Y$; $\eta_Y$ is used only as a reference for validating the re-injected texture.

To capture the dependence of relative texture amplitude on PET signal \cite{barrett1994noise}, a continuous amplitude function is fitted from the input image itself using the voxelwise pairs \((\tilde{X}(v),|\eta_X(v)|)\). Let \(s\) denote a scalar signal level. We model the conditional expected absolute texture amplitude as
\begin{equation} \mathbb{E}\left[|\eta_X| \mid \tilde{X}=s\right] = c + a \left(1+\frac{s}{s_0}\right)^{-b}, 
\label{eq:eta_abs_fit} 
\end{equation} 
with \(a>0\), \(b>0\), \(c>0\), and \(s_0>0\). This formulation reflects the empirical decrease of relative texture amplitude with increasing signal. The fitted conditional mean absolute fluctuation is then expressed on a standard deviation scale as
\begin{equation} {\sigma_X}(s) = \sqrt{\frac{\pi}{2}}\, \mathbb{E}\left[|\eta_X| \mid \tilde{X}=s\right]. 
\label{eq:sigma_hat} 
\end{equation}
This conversion is used only as a scale normalization for approximately symmetric, zero-centered fluctuations, without imposing a specific parametric noise model in the reconstructed image domain. 

At inference, the stochastic texture associated with target is not available. Therefore, MTRI uses the relative texture extracted from the input image, $\eta_X(v)$, and transfers it to the PR-corrected prediction. When PR correction changes the local signal from $\tilde{X}(v)$ to $\hat{Y}(v)$, the extracted input texture is locally rescaled according to the input-derived texture amplitude function,
\begin{equation}
A(v)=
\frac{
\sigma_X\!\left(\hat{Y}(v)\right)
}{
\sigma_X\!\left(\tilde{X}(v)\right)
}.
\end{equation}

Finally, the texture component is obtained by applying the local amplitude transfer factor to the input relative texture:
\begin{equation}
\eta_{\mathrm{MTRI}}(v)=A(v)\cdot\eta_{\mathrm{X}}(v).
\end{equation}

Importantly, $A(v)$ does not synthesize a new texture realization; it only adapts the amplitude of the texture already observed in the input image to the signal level of the PR-corrected prediction. Thus, the spatial pattern, correlations, and distributional properties of the original acquisition texture are retained, while only its local amplitude is adjusted.

\subsection{Final PRC-TP Reconstruction}
After estimating the PR-corrected structural component $\hat{Y}(v)$ and the
corrected relative texture $\eta_{\mathrm{MTRI}}(v)$, the final PRC-TP reconstructed image is obtained as
\begin{equation}
Y_{\text{PRC-TP}}(v) = \hat{Y}(v) \cdot \left[1+\eta_{\mathrm{MTRI}}(v)\right].
\label{eq:mti_output}
\end{equation}

The proposed PRC-TP framework preserves the structural correction provided by
the PRC network while reintroducing acquisition-consistent stochastic texture
extracted from the original PET image through the MTRI step.

To maintain global quantitative consistency, a renormalization step is applied:
\begin{equation}
Y_{\text{PRC-TP}} \leftarrow Y_{\text{PRC-TP}} \cdot \frac{\sum_{v \in \Omega} X(v)}{\sum_{v \in \Omega} Y_{\text{PRC-TP}}(v)},
\label{eq:mti_norm}
\end{equation}
where $\Omega$ denotes the image domain. This step compensates for global scaling changes that could be introduced by the modulation while preserving local contrast variations.

The complete PRC-TP workflow is summarized in Algorithm~\ref{alg:prc_tp}.
Starting from the input PET image and the corresponding $\mu$-map, the method first estimates the PR-corrected structural component, then extracts and signal-modulates the input texture, and finally recomposes both components multiplicatively to obtain the PR-corrected image that preserves the texture.

\begin{algorithm}[ht]
\small
\caption{Positron Range Correction with Texture Preservation (PRC-TP).}
\label{alg:prc_tp}
\begin{algorithmic}[1]

\STATE \textbf{Input:} PET image $X$, $\mu$-map, trained PRC model $\mathcal{F}_{\mathrm{PRC}}$, trained N2N model $\mathcal{F}_{\mathrm{N2N}}$
\STATE \textbf{Output:} Texture-preserving PR-corrected image $Y_{\mathrm{PRC\text{-}TP}}$

\vspace{0.2em}
\STATE \textbf{Step 1: Positron range correction}
\STATE $\hat{Y} = \mathcal{F}_{\mathrm{PRC}}(X,\mu)$

\vspace{0.2em}
\STATE \textbf{Step 2: N2N signal estimation}
\STATE $\tilde{X} = \mathcal{F}_{\mathrm{N2N}}(X,\mu)$

\vspace{0.2em}
\STATE \textbf{Step 3: Relative texture extraction}
\STATE $\eta_X(v) = X(v)/\tilde{X}(v) - 1$

\vspace{0.2em}
\STATE \textbf{Step 4: Signal-dependent texture-amplitude fitting}
\STATE \hspace{0.5cm}\textbf{Fit} the conditional mean absolute fluctuation
\[
\mathbb{E}\left(|\eta_X|\,\middle|\,\tilde{X}=s\right)
=
c+a\left(1+\frac{s}{s_0}\right)^{-b}
\]
\STATE \hspace{0.5cm}\textbf{using} voxelwise pairs
$\left(\tilde{X}(v),|\eta_X(v)|\right)$
\STATE \hspace{0.5cm}\textbf{with} $a>0$, $b>0$, $c>0$, and $s_0>0$
\STATE \hspace{0.5cm}
$\sigma_X(s) = \sqrt{\pi/2}\,
\mathbb{E}\left(|\eta_X|\,\middle|\,\tilde{X}=s\right)$
\vspace{0.2em}
\STATE \textbf{Step 5: Signal-dependent texture modulation}
\STATE $A(v) =
\dfrac{
\sigma_X\left(\hat{Y}(v)\right)}
{\sigma_X\left(\tilde{X}(v)\right)}$
\STATE $\eta_{\mathrm{MTRI}}(v) = A(v)\,\eta_X(v)$

\vspace{0.2em}
\STATE \textbf{Step 6: Multiplicative recomposition}
\STATE $Y_{\mathrm{PRC\text{-}TP}}(v) =
\hat{Y}(v)\left[1+\eta_{\mathrm{MTRI}}(v)\right]$

\vspace{0.2em}
\STATE \textbf{return} $Y_{\mathrm{PRC\text{-}TP}}$

\end{algorithmic}
\end{algorithm}

\section{Materials and Methods}
\subsection{Patient-Derived Datasets}
A simulation-based patient-derived cohort was used for supervised training and quantitative evaluation.  All patient data used in this study were acquired at Washington University (St. Louis, MO, USA). The study was approved by the Washington University Institutional Review Board (IRB) and all patients provided written informed consent.

The synthetic cohort was derived from 38 cardiac patients originally imaged with \textsuperscript{13}N-ammonia on a Biograph Vision PET/CT system (Siemens Healthineers, Knoxville, TN, USA). \textsuperscript{13}N-ammonia was selected to provide a PET-derived baseline with small PR blurring, enabling the construction of patient-specific voxelized anthropomorphic phantoms for Monte Carlo simulations. The reconstructed \textsuperscript{13}N volumes were used as input activity distributions for GATE  \cite{jan2004gate}simulations emulating the emission physics of \textsuperscript{82}Rb. For supervised training, corresponding back-to-back (b2b) simulations were generated assuming annihilation photons emitted from the same spatial location as the positron emission and an ideal collinearity of the resulting 511~keV photons. These b2b volumes represent ideal emissions without PR or non-collinearity effects and served as the GT. The resulting synthetic dataset was partitioned into training (n=25), validation (n=9), and independent testing (n=4) subsets.

\vspace{-0.15cm}
\subsection{Monte Carlo Simulation and Reconstruction}
Monte Carlo simulations were performed using GATE \cite{jan2004gate, sarrut2021advanced, sarrut2022opengate} (v9.1) built on Geant4 \cite{collaboration2003geant4, allison2006geant4, allison2016recent} (v10.7.2). A geometrical model of the Biograph Vision PET/CT scanner (hereafter Vision) was implemented to reproduce the specifications of the commercial system. The design included a bore diameter of 82 cm (crystal-to-crystal) and it consisted of a single cylindrical segment with an axial field of view (aFOV) of 26.1 cm and 760 detectors per ring, organized in blocks of 80 lutetium oxyorthosilicate (LSO) crystals with dimensions of $3.2 \times 3.2 \times 20$~mm$^3$.

The electromagnetic interactions were simulated using the \textit{emstandard$\_$opt4} physics list in GATE \cite{beaudoux2019geant4, salvadori2020monte}. Deposited energies within each LSO crystal were aggregated into single-crystal events and then combined into detector-level events. To reproduce the intrinsic detector response, a 9\% energy resolution was applied, and the resulting data were exported in ROOT format \cite{brun1997root} without applying an energy cut, thereby preserving low-energy contributions. Event processing was subsequently performed with the investigational root2lm tool (Siemens Healthineers, Knoxville, TN, USA), which models system-specific detector characteristics such as timing resolution, dead time, intercrystal scatter, and intrinsic LSO background, while applying the same coincidence-sorting algorithm as in the clinical scanners. Unlike generic dead-time or pile-up corrections often implemented in GATE workflows \cite{salvadori2024pet}, root2lm incorporates vendor-specific component models, providing a bottom-up representation of the hardware. The output consists of PET listmode files in the PETLINK format \cite{jones2013petlink, jones2018pet}, fully consistent with raw Vision acquisitions and including prompts, randoms, and acquisition tags (time, singles, bed position).

Image reconstructions were performed with the investigational e7 tools software (Siemens Healthineers, Knoxville, TN, USA), using the same processing pipeline as the clinical systems. This included histogramming of the listmode data, attenuation and scatter correction, normalization, decay correction, and calibration to activity concentration. Attenuation maps were generated within GATE using the MuMapActor, with an isotropic voxel size of 1.65 mm. For normalization, a component-based file \cite{casey1995component} derived from the physical scanners was applied, with calibration factors adapted to the simulated acquisitions. All simulations presented in this work were executed in a computer farm consisting of 32 HP ZCentral 4R workstations Intel(R) Xeon(R) W-2295 CPU @ 3.00GHz with 18 cores (two threads/core).

\vspace{-0.15cm}
\subsection{Model Architecture and Training Strategy}
The PRC and N2N model are based on the nnFormer architecture \cite{zhou2021nnformer}, a hybrid model that integrates Swin Transformer blocks into a 3D U-Net backbone \cite{ronneberger2015u}. This design synergizes the local feature extraction capabilities of CNNs \cite{lecun1998convolutional} with the long range dependency modeling of Vision Transformers \cite{shamshad2023transformers}, allowing the model to effectively contextualize metabolic intensities within their anatomical surroundings. The network operates on a dual channel input (PET and CT-derived $\mu$-map), fusing functional data with anatomically derived spatial information to guide the training process. 

To address the local nature of PR effects and reduce the risk of the network overfitting to global tracer distribution patterns, the model was trained using volumetric patches ($64 \times 64 \times 64$ voxels). The patch size was chosen to be sufficiently large to capture local tissue interfaces affected by PR while remaining computationally feasible for 3D transformer-based training. We implemented a gradient-smart patch sampling strategy to focus on regions where PR blurring is most critical. To this end, 3D patches were sampled based on ground-truth gradient maps, prioritizing edges and tissue boundaries where mean gradient values exceeded the 95th percentile of the volume. To ensure model robustness and prevent overfitting to edges, a stochastic fraction ($2\%$) of low-gradient patches was retained to preserve coverage of homogeneous regions and prevent structural bias during optimization. The percentile threshold and uniform sampling fraction were empirically selected to balance emphasis on tissue interfaces while preserving sufficient coverage of homogeneous regions.

\vspace{-0.15cm}
\subsection{Objective Function and Adaptive Optimization}
The optimization framework builds upon the tissue-dependent loss formulation proposed in our previous work \cite{encina2026tissue}. The same objective function was used to train both the PRC and N2N models, ensuring that differences between both stages arise from the training pairs rather than from the optimization criterion. 

To keep the notation general, let $P$ denote the network prediction and $T$ the corresponding training target. For PRC, $(P,T) = (\hat{Y}, Y)$, whereas for N2N $(P,T)=(\tilde{X}, X')$.  The loss function is defined as
\begin{equation}
\mathcal{L}(P, T, \mu)
=
\mathcal{L}_{\mathrm{MAE}}(P, T)
+
\gamma_t \, \mathcal{L}_{\mathrm{Reg}}(P, T, \mu),
\label{eq:loss_general}
\end{equation}
where $\mu$ denotes the attenuation map. The term
$\mathcal{L}_{\mathrm{MAE}}$ enforces voxel-wise activity agreement, while the regularization term promotes anatomical consistency through Mutual Information (MI):
\begin{equation}
\mathcal{L}_{\mathrm{Reg}}(P, T, \mu)
=
\left|
\mathrm{MI}(T,\mu)
-
\mathrm{MI}(P,\mu)
\right|.
\label{eq:mi_regularization_general}
\end{equation}

The relative contribution of this regularization term is controlled by the dynamic factor $\gamma_t$. 

In this work, two key extensions are introduced to improve robustness and practical usability: (i) a residual-guided spatial reweighting strategy that yields a weighted MAE and weighted MI formulation, and (ii) an adaptive hyperparameter balancing scheme that removes the need for manual tuning of the regularization weight.

\subsubsection{Residual-Guided Spatial Reweighting}
To account for the spatially heterogeneous impact of local image degradation,
we introduce a residual-guided reweighting strategy during training. A voxel-wise residual map is computed for each
training pair as
\begin{equation}
d(v)
=
\left(
G \ast |X-T|
\right)(v),
\label{eq:residual_map}
\end{equation}
where $G$ is a fixed $3 \times 3 \times 3$ Gaussian kernel. The Gaussian
smoothing suppresses isolated fluctuations and yields a spatially coherent estimate of local discrepancies between input $X$ and target $T$.

The residual map is then normalized within each training patch using a
spatial Softmax:
\begin{equation}
w(v)
=
\frac{
\exp(d(v))
}{
\sum_{v' \in \Omega} \exp(d(v'))
},
\label{eq:residual_softmax}
\end{equation}
where $\Omega$ denotes the spatial domain of the patch. After normalization, $w(v)$ is a non-negative weighting map satisfying
$\sum_{v \in \Omega}w(v)=1$, which increases the relative contribution of voxels with larger local discrepancy.

The weighting map is used only during loss evaluation and does not modify the network inputs or outputs. Specifically, the prediction, target, and attenuation map are weighted before evaluating the MAE and MI terms:
\begin{equation}
\mathcal{L}
=
\mathcal{L}_{\mathrm{MAE}}(w\odot P,\,w\odot T)
+
\gamma_t\,
\mathcal{L}_{\mathrm{Reg}}(w\odot P,\,w\odot T,\,w\odot\mu),
\label{eq:weighted_loss_general}
\end{equation}
where $\odot$ denotes voxel-wise multiplication.

\subsubsection{Adaptive Hyperparameter Balancing}

To reduce reliance on manually selecting the regularization weight, we introduced an adaptive hyperparameter balancing strategy that dynamically adjusts $\gamma_t$. We defined a target ratio $\rho = 0.1$, representing the desired contribution of the regularization term relative to the reconstruction loss (i.e., $\gamma \cdot \mathcal{L}_{\mathrm{Reg}} \approx \rho \cdot \mathcal{L}_{\mathrm{MAE}}$), ensuring that the optimization remains primarily driven by PET reconstruction fidelity while the anatomical term acts as a secondary structural constraint.

At the end of each epoch, a candidate $\gamma_{\text{new}}$ was computed based on the ratio of the accumulated loss components. To prevent oscillations and ensure smooth convergence, the update follows an Exponential Moving Average (EMA):
\begin{equation}
\gamma_{t} = \beta \gamma_{t-1} + (1 - \beta) \gamma_{\text{new}}, \quad 
\gamma_{\text{new}} = \rho \cdot \frac{\mathcal{L}_{\mathrm{MAE}}}{\mathcal{L}_{\mathrm{Reg}}}
\end{equation}
where $\beta = 0.9$ is the momentum factor. The value was constrained to $\gamma_t \in [0.05, 1.0]$. In our experiments, $\gamma$ was initialized at $0.3$ and converged to a stable value near $0.2$, effectively maintaining the trade-off between structural guidance and quantitative accuracy without manual tuning.

\vspace{-0.15cm}
\subsection{Implementation Details}
\subsubsection{Data Processing and Augmentation}

To ensure numerical stability and suppress reconstruction artifacts, PET
intensities were clipped above the 99.99th percentile to remove isolated hot
voxels. A patient-specific scaling factor, defined as the mean intensity of
the input PET volume, was used to normalize both $^{82}$Rb and b2b images.
Attenuation maps ($\mu$-maps) were scaled by a factor of 50 to match the
dynamic range of the normalized PET images. This scaling was applied solely
for numerical conditioning and does not affect their physical interpretation.

Normalization was performed at the full-volume level to preserve inter-tissue
intensity relationships. Patch-wise normalization was intentionally avoided,
as it would equalize local activity differences and could implicitly impose
local conservation. All processed volumes, gradient maps, and valid patch
coordinates were stored in HDF5 format for efficient data loading.

Data augmentation during training was limited to random orthogonal flips.
Spatial rescaling and intensity perturbations were excluded to preserve the
physical relationship between PR effects and the underlying activity
distribution.

\subsubsection{Computational Setup and Optimization}

All experiments were implemented in PyTorch with CUDA 13.0 and conducted on
a workstation equipped with an NVIDIA GeForce RTX 5090 GPU with 32~GB of
GDDR7 VRAM. The models were trained using input patches of
$64\times64\times64$ voxels. To avoid boundary artifacts caused by
zero-padding in convolutional layers, the loss was computed only over the
central $56\times56\times56$ region of each patch.

The networks were optimized with AdamW using a weight decay of
$1\times10^{-3}$. A composite learning-rate schedule was used, consisting of a linear warmup to a peak learning rate of 5×10$^{-4}$ (75 epochs for PRC and 15 epochs for N2N), followed by cosine annealing. A batch size of 42 was used, and early stopping
was applied after 100 epochs without validation improvement. Each epoch included 10,000 training patches and 4,000 validation patches.

\subsubsection{Inference and Model Selection}

Inference was performed using a grid-based sampling strategy implemented with
the TorchIO library \cite{perez2021torchio}. The models predicted patches of $64^3$ voxels, but only
the central $56^3$ effective field of view was retained to avoid boundary
artifacts introduced by convolutional padding. Whole-volume predictions were obtained by combining overlapping patch predictions using weighted averaging, giving higher weight to central patch regions to reduce stitching and boundary artifacts.

Model selection was based on the validation loss. The best-performing
checkpoints were obtained at epochs 116 and 54 for the PRC and N2N models,
respectively.
\vspace{-0.3cm}
\subsection{Evaluation}
The proposed framework was assessed on simulated patients, for which GT images were available, to assess texture consistency, radiomics preservation, PR
correction, and quantitative accuracy. The texture re-injection step was first evaluated independently, since the objective was not to
reproduce the GT texture voxelwise, but to recover its statistical properties. We therefore compared the signal-dependent texture amplitude of
the input-derived and GT-derived textures as a function of the corresponding smoothed signal, and also evaluated whether the re-injected texture remained coupled to local PR-related signal changes. For this purpose, we defined a PR-change proxy as
\begin{equation}
\Delta_{PR}(v)=
\frac{\hat{Y}(v)-\tilde{X}(v)}
{\tilde{X}(v)}.
\label{eq:delta_r}
\end{equation}

Radiomics analysis was used to assess whether PRC-TP preserve texture-dependent image features. Features were extracted from myocardium, left ventricle, liver, and lung ROIs and grouped into first-order (FO), gray-level co-occurrence matrix (GLCM), gray-level run-length matrix (GLRLM), and gray-level size-zone matrix (GLSZM) families, describing intensity statistics, spatial texture relationships, run-length patterns, and homogeneous zone-size distributions, respectively. Each method was compared with GT using a normalized radiomics distance, where lower values indicate closer agreement.

Image quality and quantitative recovery were evaluated using residual maps, line profiles, and functional PET metrics. Recovery coefficient (RC), contrast-to-noise ratio (CNR), and liver noise were computed to assess contrast recovery, contrast detectability, and noise preservation, respectively. The contrast ratio (CR) was defined as the mean myocardial activity divided by the mean left-ventricular activity, and RC was obtained by normalizing CR to the corresponding GT value. Liver noise was defined as the liver coefficient of variation, whereas CNR was computed as the myocardium-to-left-ventricle contrast divided by the liver standard deviation. For all patients, left ventricular myocardium and cavity segmentations were generated in ITK-SNAP 4.4.0 \cite{itk} using semi-automatic active contours and were manually reviewed and refined. The liver ROI was manually delineated.

To assess the practical consistency of the proposed framework beyond the simulation environment, PRC-TP was also applied to a real clinical $^{82}$Rb PET study. Since a ground-truth reference was not available in that case, the analysis was limited to reference-free image quality and functional metrics. The objective was not to establish quantitative accuracy, but rather to evaluate whether the trends observed in simulated patients, including contrast recovery, noise restoration, and texture preservation, remained consistent in real clinical data.
\section{Results}
\subsection{Validation of Texture Transfer Consistency}
\begin{figure*}
    \centering
    \includegraphics[width=1\linewidth]{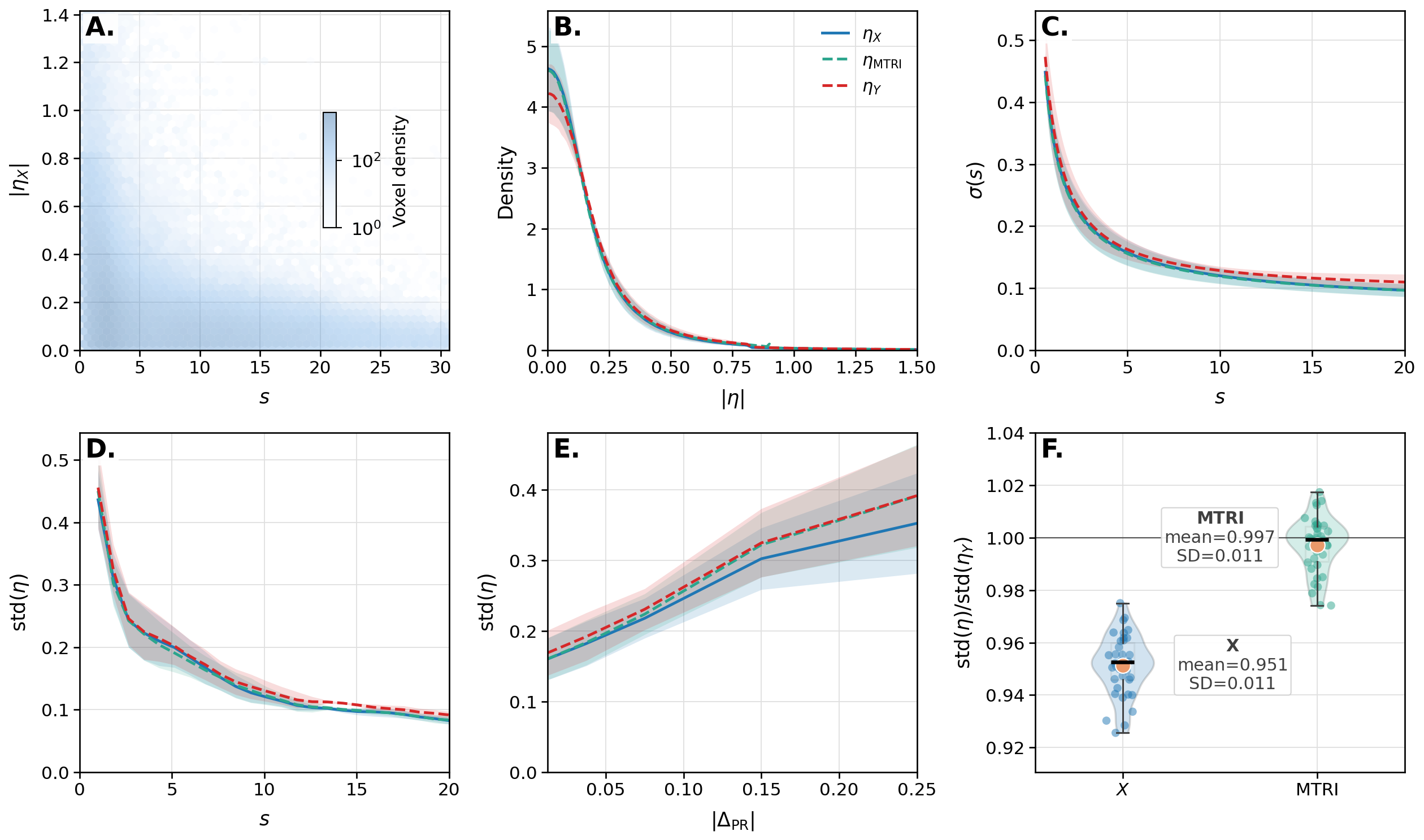}
\caption{
Validation of the MTRI texture model in the simulation cohort. Here, \(X\) denotes the PR-degraded input image and \(Y\) denotes the PR-free ground truth (GT).
\textbf{(A)} Voxelwise density map of the absolute texture amplitude extracted from \(X\), \(|\eta_X|\), as a function of signal level \(s\), illustrating the signal-dependent decrease in relative texture amplitude.
\textbf{(B)} Distributions of absolute texture amplitude for \(X\), MTRI, and \(Y\) textures, illustrating statistical agreement rather than voxelwise matching.
\textbf{(C)} Signal-dependent texture amplitude functions, expressed as \(\sigma(s)\), for \(X\), MTRI, and \(Y\).
\textbf{(D)} Empirical texture standard deviation as a function of signal level \(s\), showing the agreement between MTRI and \(Y\) across signal levels.
\textbf{(E)} Texture standard deviation as a function of the PR-change proxy \(|\Delta_{\mathrm{PR}}|\), used to assess whether the re-injected texture remains consistent across regions with different PR-related signal changes.
\textbf{(F)} Global texture amplitude ratio relative to \(Y\), showing that MTRI reduces the texture amplitude bias of \(X\) and brings the global texture scale close to \(Y\).
Panels A--D were computed on the simulated test cohort, whereas panels E--F were computed using all 38 simulated patients to improve the stability of the cohort-level bias and PR-change analyses.
The same colors for \(X\), MTRI, and \(Y\) are used across panels B--F and are indicated in the legend of panel B.
}
    \label{fig:texture_injection_consistency}
\end{figure*}

Figure~\ref{fig:texture_injection_consistency} evaluates the cohort-level consistency of the function-based MTRI texture model. The validation was performed progressively: first, by characterizing the signal-dependent relationship between the absolute input texture amplitude, $|\eta_X|$, and the local signal $s$, then, by comparing the distributions and signal-dependent texture amplitude functions of the input, MTRI, and $Y$ (GT) textures, and finally, by assessing whether MTRI reduces the global texture amplitude bias relative to GT and remains consistent across different levels of PR-related signal change.

Figures~\ref{fig:texture_injection_consistency}A--C characterize the signal-dependent behavior of the proposed texture model. The absolute input texture amplitude decreases with increasing signal, confirming the heteroscedastic nature of reconstructed PET texture and supporting the use of a signal-dependent amplitude function. The texture amplitude distributions show that both input and MTRI textures closely follow the $Y$ distribution. Furthermore, the fitted texture amplitude functions indicate that the input-derived model provides a close approximation to the GT relationship across the full signal range. Figure~\ref{fig:texture_injection_consistency}D shows the empirical texture standard deviation as a function of signal. Consistent with the fitted models, MTRI reproduces the signal-dependent texture amplitude observed in $Y$ across the entire signal range. This analysis is particularly relevant because the standard deviation directly quantifies the amplitude of stochastic fluctuations around the local signal, independently of their voxelwise realization. Therefore, the observed agreement between $X$, MTRI, and $Y$ demonstrates that the proposed transfer preserves the correct signal-dependent texture scale rather than merely reproducing a similar overall texture distribution.

The dependence of texture amplitude on the PR-change proxy $|\Delta_{PR}|$ is shown in Figure~\ref{fig:texture_injection_consistency}E. This analysis complements the signal-based evaluation by focusing on regions where PR correction produces larger local signal changes and where texture rescaling is therefore most relevant. In these regions, the input texture underestimates the GT amplitude, whereas MTRI remains closely aligned with GT across the range of PR-related changes. Finally, Figure~\ref{fig:texture_injection_consistency}F summarizes this effect globally. The input texture shows an amplitude ratio of $0.951\pm0.011$ relative to $Y$, whereas MTRI increases this value to $0.997\pm0.011$, confirming near-unity texture amplitude agreement after signal-dependent transfer.

\begin{figure*}
    \centering
    \includegraphics[width=1\linewidth]{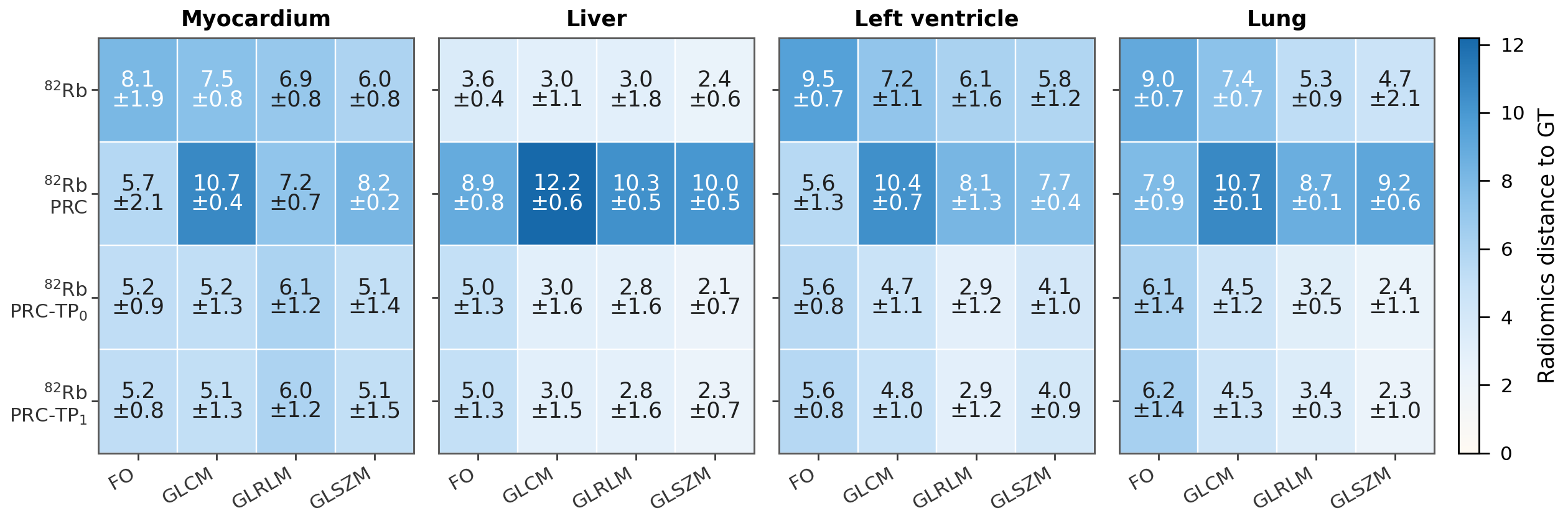}
\caption{Radiomics agreement with ground truth (GT) across anatomical ROIs. Heatmaps show the mean radiomics distance to GT for each method and feature family in the myocardium, liver, left ventricle, and lung. Feature families include first-order (FO), gray-level co-occurrence matrix (GLCM), gray-level run-length matrix (GLRLM), and gray-level size-zone matrix (GLSZM) descriptors. Each cell reports the mean $\pm$ SD across patients, with lower values indicating closer agreement with GT. PRC-TP$_0$ corresponds to texture re-injection without signal-dependent amplitude modulation ($A(v)=1$), whereas PRC-TP$_1$ includes the proposed signal-dependent modulation factor $A(v)$. Lighter colors indicate lower radiomics distance and therefore closer agreement with GT.}

    \label{fig:radiomics}
\end{figure*}
\vspace{-0.15cm}
\subsection{Radiomics Analysis}

\begin{figure*}
    \centering
    \includegraphics[width=1\linewidth]{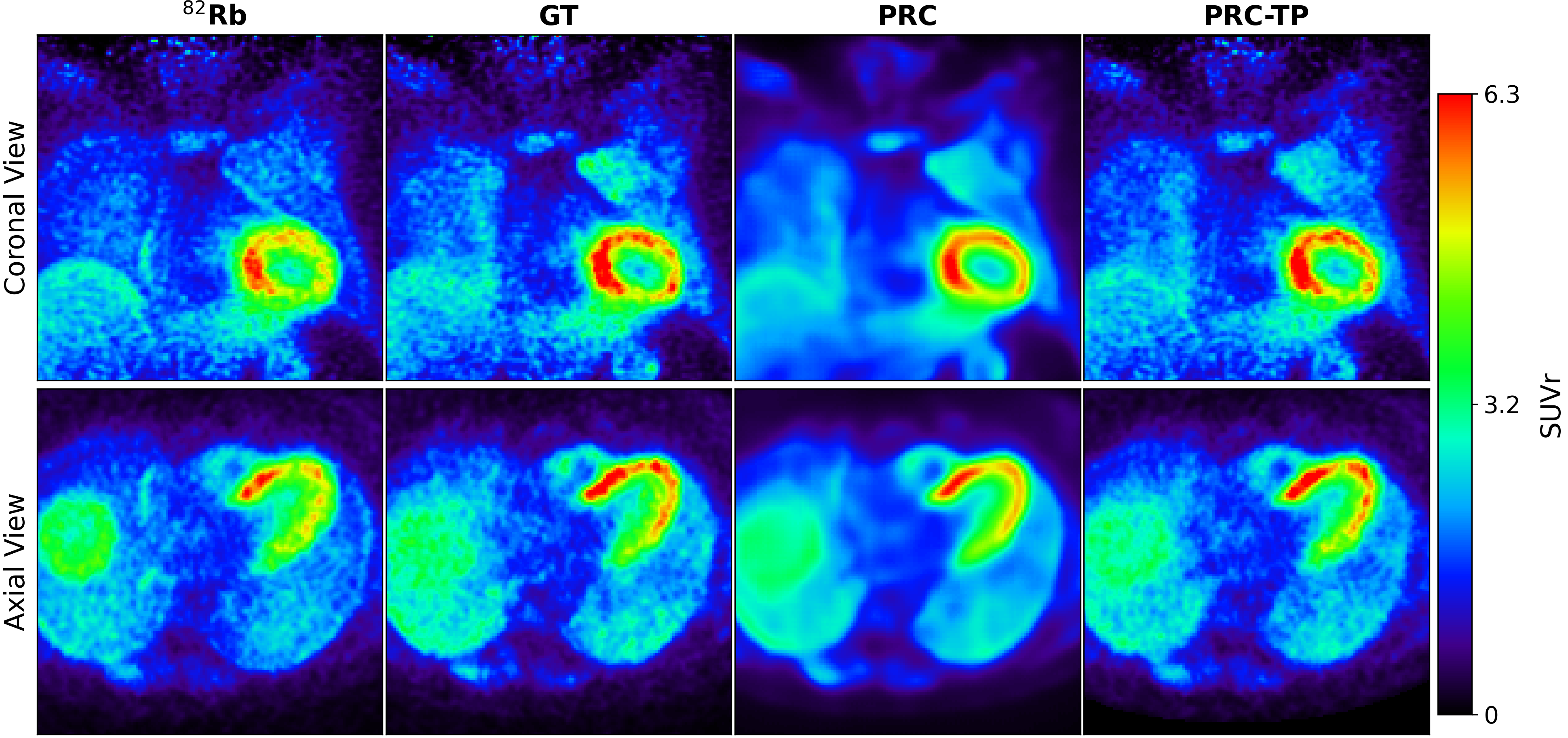}
    \caption{Representative qualitative comparison for one patient of the test cohort. Coronal and axial views are shown for the input $^{82}\mathrm{Rb}$ PET image, GT, PRC prediction, and PRC-TP reconstruction. PRC-TP preserves the PR-corrected anatomical structure recovered by PRC while reintroducing image texture consistent with the input acquisition. Images are displayed in SUVr using the same intensity scale within each view.}
    \label{fig:patient}
\end{figure*}

Figure~\ref{fig:radiomics} summarizes the radiomics agreement with GT across anatomical ROIs and feature families. PRC-TP$_0$ denotes texture re-injection without signal-dependent amplitude modulation ($A(v)=1$), whereas PRC-TP$_1$ includes the proposed modulation factor $A(v)$. The PRC output showed increased radiomics distance in several texture-sensitive families, particularly GLCM, GLRLM, and GLSZM, consistent with the attenuation of stochastic texture in the smooth PRC prediction. In contrast, both PRC-TP variants reduced the distance to GT across most ROIs and feature families, with the largest improvements observed in the left ventricle and lung texture features.

The improvement was mainly observed relative to PRC, whereas PRC-TP$_0$ and PRC-TP$_1$ showed similar performance. This suggests that radiomics preservation was primarily driven by MTRI texture re-injection, with limited additional effect from signal-dependent amplitude modulation.
\begin{table*}[t]
\centering
\caption{Functional PET metrics across patients for the proposed PRC-TP variants. Values are reported as mean $\pm$ SD across patients. PRC-TP$_0$ corresponds to direct texture re-injection without signal-dependent amplitude modulation, whereas PRC-TP$_1$ includes the proposed signal-dependent modulation factor $A(v)$.}
\label{tab:functional_prctp_results}
\renewcommand{\arraystretch}{1.15}
\resizebox{\textwidth}{!}{%
\begin{tabular}{lcccccc}
\toprule
\multirow{2}{*}{\textbf{Method}} 
& \multicolumn{3}{c}{\textbf{Metrics}} 
& \multicolumn{3}{c}{\textbf{Relative error to GT (\%)}} \\
\cmidrule(lr){2-4} \cmidrule(lr){5-7}
& \textbf{RC (\%)} 
& \textbf{CNR} 
& \textbf{Noise (\%)} 
& \textbf{$\Delta$RC} 
& \textbf{$\Delta$CNR} 
& \textbf{$\Delta$Noise} \\
\midrule
\textbf{GT $\left(^{13} \text{N}\right)$} 
& $100.00 \pm 0.00$ 
& $7.02 \pm 0.86$ 
& $10.62 \pm 0.82$ 
& -- 
& -- 
& -- \\

\textbf{$^{82}$Rb} 
& $89.12 \pm 3.60$ 
& $7.12 \pm 1.49$ 
& $9.07 \pm 1.70$ 
& $-10.88 \pm 3.60$ 
& $2.13 \pm 24.18$ 
& $-14.09 \pm 18.42$ \\

\textbf{PRC} 
& $96.76 \pm 1.07$ 
& $12.82 \pm 4.64$ 
& $5.98 \pm 1.91$ 
& $-3.24 \pm 1.07$ 
& $83.08 \pm 62.57$ 
& $-43.48 \pm 17.80$ \\

\textbf{PRC-TP$_0$} 
& $98.96 \pm 1.75$ 
& $6.96 \pm 1.05$ 
& $10.83 \pm 1.04$ 
& $-1.04 \pm 1.75$ 
& $\mathbf{-0.85 \pm 9.68}$ 
& $\mathbf{2.22 \pm 9.99}$ \\

\textbf{PRC-TP$_1$} 
& $99.04 \pm 1.76$ 
& $6.94 \pm 1.04$ 
& $10.86 \pm 1.02$ 
& $\mathbf{-0.96 \pm 1.76}$ 
& $-1.08 \pm 9.39$ 
& $2.51 \pm 9.78$ \\
\bottomrule
\end{tabular}%
}
\end{table*}

\begin{figure}
    \centering
    \includegraphics[width=0.96\linewidth]{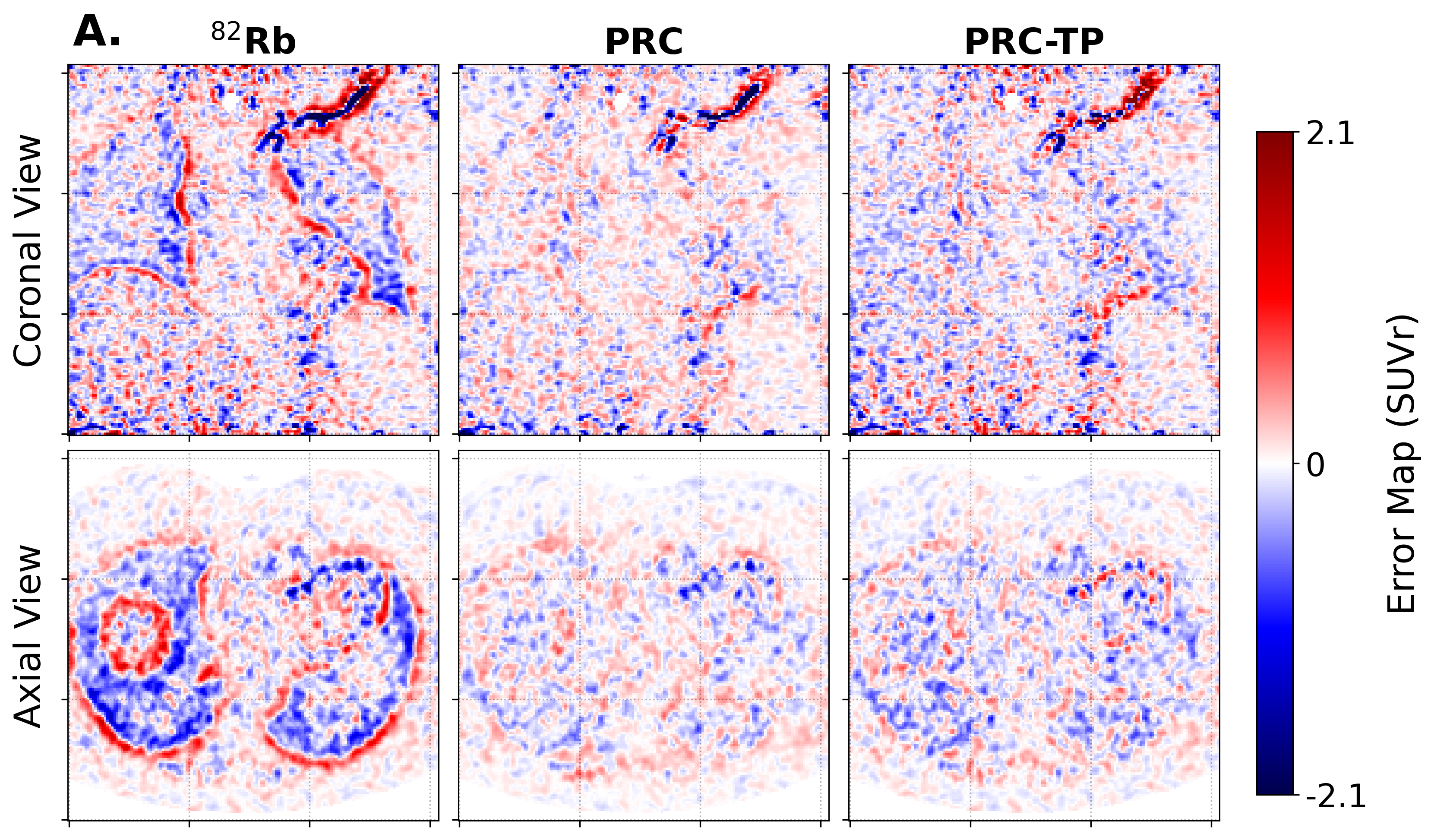}

    \includegraphics[width=0.99\linewidth]{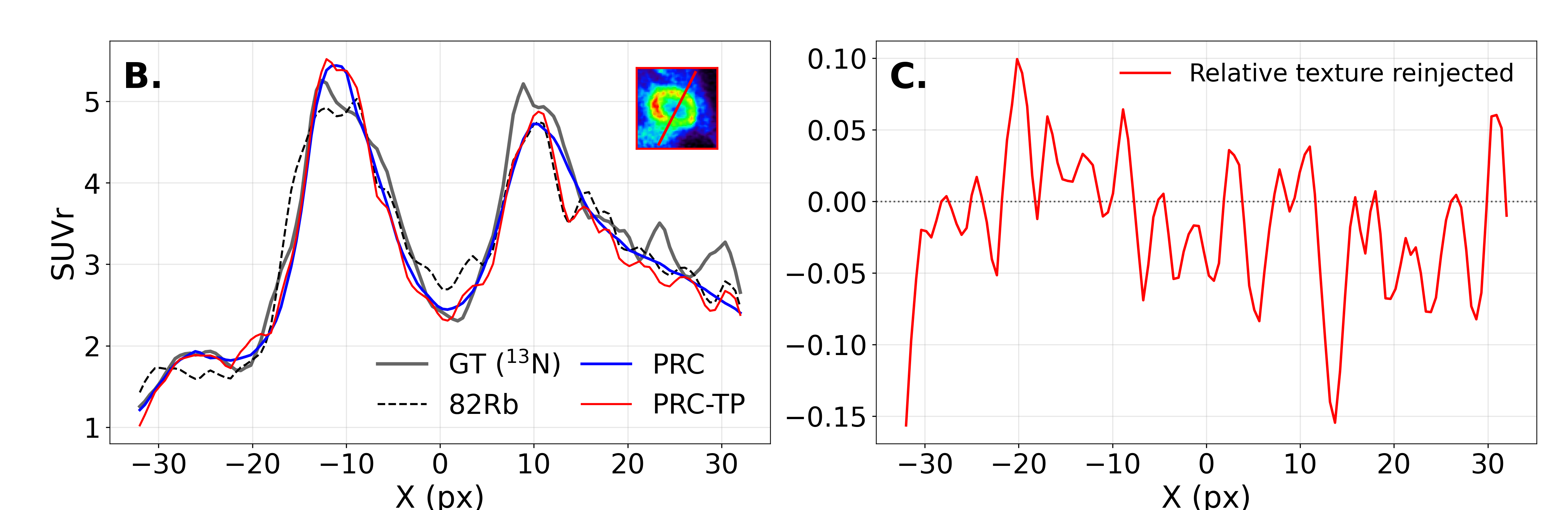}
    \caption{Residual and profile analysis for a representative patient. \textbf{A)} error maps computed with respect to GT for the input $^{82}\mathrm{Rb}$ PET image, PRC prediction, and PRC-TP reconstruction in coronal and axial views. Residuals are shown only within the body mask using a common symmetric color scale. \textbf{B)} coronal intensity profile comparing GT, input, PRC, and PRC-TP along the indicated line. \textbf{C)} relative texture reinjected into the PRC prediction, computed as $((\mathrm{PRC\text{-}TP})-\mathrm{PRC})/\mathrm{PRC}$, along the same coronal profile.}
    \label{fig:error}
\end{figure}

\vspace{-0.25cm}
\subsection{Quantitative Evaluation on Simulated Patients}
Figure~\ref{fig:patient} shows a representative qualitative comparison of the input $^{82}$Rb PET image, GT ($^{13}$N), PRC prediction, and final PRC-TP reconstruction. PRC reduced PR blurring and recovered sharper anatomical structures, but produced a visually smoother image due to attenuation of stochastic PET texture. In contrast, PRC-TP preserved the anatomical signal recovered by PRC while restoring an acquisition consistent texture pattern closer to GT.

Residual maps and line profiles are shown in Fig.~\ref{fig:error}. The input $^{82}$Rb image exhibits structured deviations relative to GT, consistent with PR spill out and contrast loss. PRC reduced this structural bias and recovered a sharper activity profile, but also suppressed high frequency stochastic texture. PRC-TP preserved the activity profile recovered by PRC while reintroducing fluctuations consistent with the original acquisition. The line profiles in Fig.~\ref{fig:error}B--C further show that PRC-TP does not reintroduce the PR blurring corrected by PRC, but rather adds local stochastic fluctuations around the activity profile recovered by PRC.

Table~\ref{tab:functional_prctp_results} summarizes the functional PET metrics across patients. The input $^{82}$Rb image underestimated contrast recovery, with an RC of $89.12 \pm 3.60\%$, whereas PRC improved RC to $96.76 \pm 1.07\%$. However, this improvement was accompanied by a marked reduction in liver noise relative to GT ($-43.48 \pm 17.80\%$), which led to an overestimation of CNR ($83.08 \pm 62.57\%$).

Both PRC-TP variants preserved the contrast recovery achieved by PRC while restoring noise and CNR closer to the GT reference. PRC-TP$_0$ achieved the best CNR and noise agreement, with errors of $-0.85 \pm 9.68\%$ and $2.22 \pm 9.99\%$, respectively. PRC-TP$_1$ provided the closest RC agreement ($-0.96 \pm 1.76\%$) and showed similar noise restoration. Overall, these results indicate that PRC-TP maintains the deterministic resolution recovery of PRC while recovering acquisition-consistent texture and noise levels.
\vspace{-0.15cm}
\subsection{Evaluation on Real Data}
\begin{figure}
    \centering
    \includegraphics[width=0.96\linewidth]{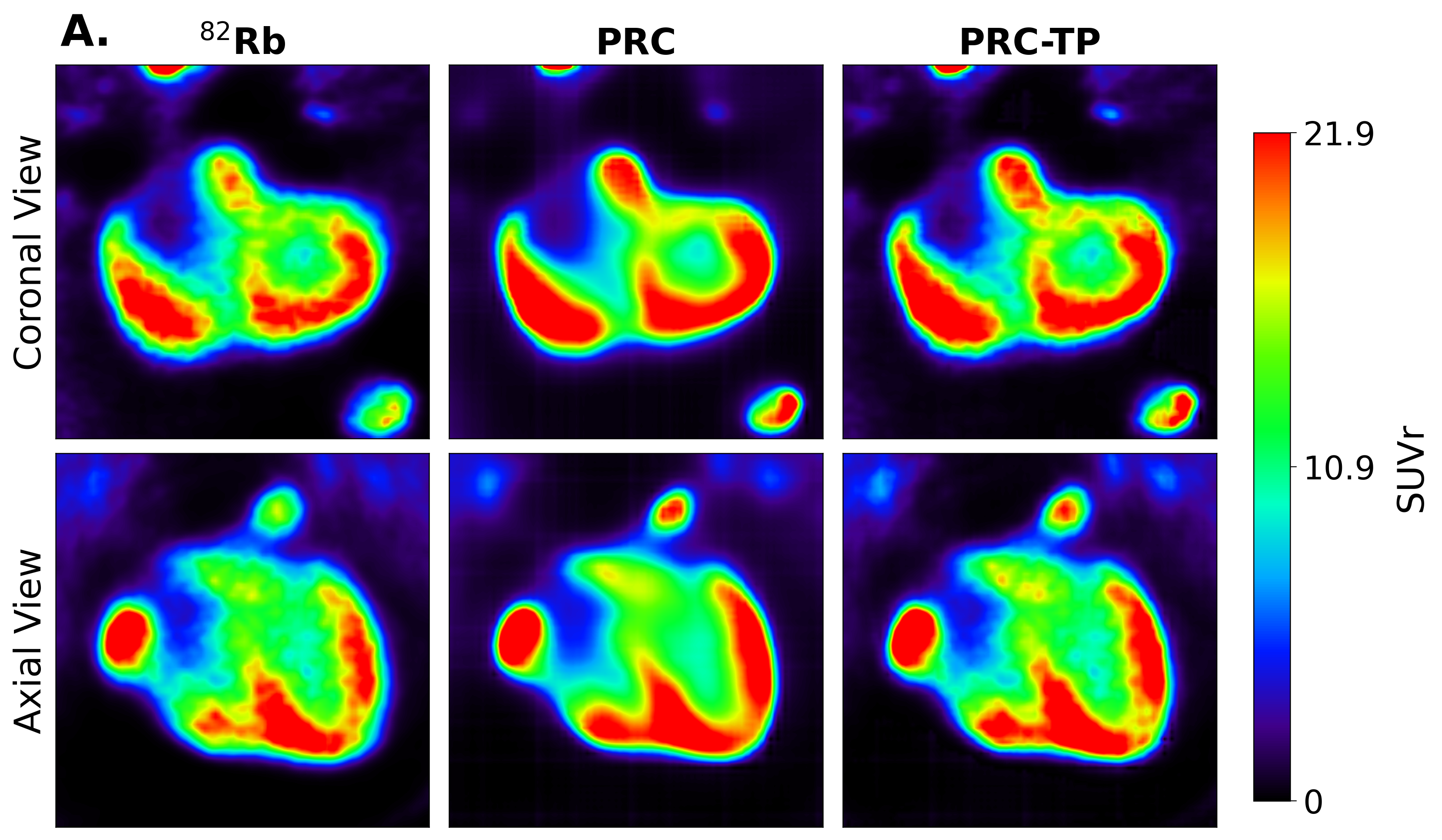}

    \includegraphics[width=0.99\linewidth]{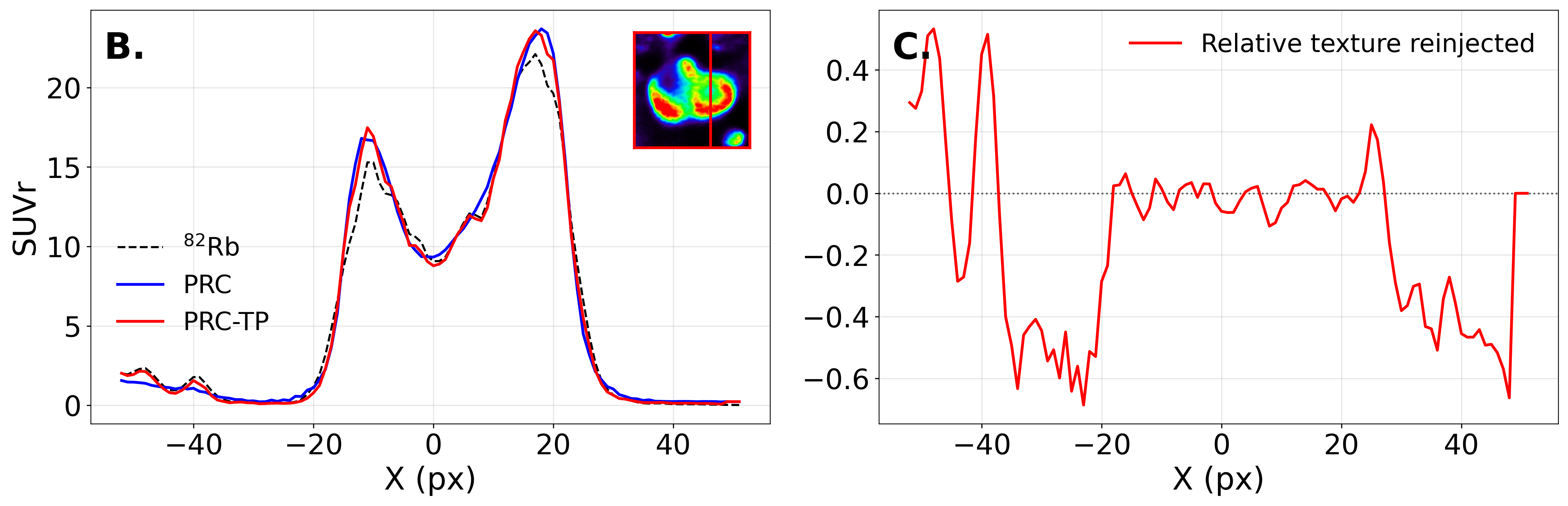}
\caption{
Representative clinical $^{82}$Rb PET example illustrating the effect of PRC-TP.
\textbf{(A)} Coronal and axial views of the original $^{82}$Rb image, the PRC prediction, and the final PRC-TP reconstruction. PRC recovers sharper myocardial structures but attenuates stochastic PET texture, whereas PRC-TP restores acquisition-consistent texture while preserving the PR-corrected anatomical signal.
\textbf{(B)} Line profile across the myocardium (location indicated in the inset). PRC-TP closely follows the PRC activity profile, indicating that texture re-injection does not reintroduce the blurring corrected by PRC.
\textbf{(C)} Relative texture modulation re-injected by MTRI along the same profile, computed as $(\mathrm{PRC\text{-}TP}-\mathrm{PRC})/\mathrm{PRC}$. The re-injected component behaves as a local relative fluctuation around the PRC prediction rather than as an additive structural correction.
All images are displayed in SUVr using a common intensity scale.
}
    \label{fig:pat}
\end{figure}

Figure~\ref{fig:pat} shows a representative qualitative comparison between the original $^{82}$Rb image, the PRC prediction, and the final PRC-TP reconstruction in a real clinical study. PRC recovered sharper myocardial structures and reduced the apparent PR blurring present in the original image, but also attenuated the stochastic texture characteristic of PET acquisitions. PRC-TP restored this texture while preserving the anatomical profile recovered by PRC.

The corresponding line profiles are shown in Fig.~\ref{fig:pat}B--C. The close agreement between the PRC and PRC-TP profiles confirms that texture re-injection preserves the spatial resolution recovered by PRC. The re-injected component acts as a local relative modulation around the PRC prediction, producing fluctuations consistent with the original acquisition without altering the corrected anatomical profile.
\section{Discussion}
In this work, we proposed the PRC-TP framework that explicitly separates structural resolution recovery from stochastic texture preservation in PET. Although evaluated here in $^{82}$Rb cardiac PET, the formulation is defined in the image domain and is not intrinsically restricted to a specific radionuclide. The PRC network recovers the deterministic structural component affected by PR, whereas the N2N model estimates a matched smoothed representation of the input. This enables MTRI to restore texture consistent with the acquired PET image in a controlled and deterministic manner, rather than requiring the PRC network to preserve stochastic fluctuations that are intrinsically unpredictable. The results show that PRC improves contrast recovery but suppresses image texture and noise variability, as expected for supervised regression models trained to estimate a deterministic target from noisy data. PRC-TP addresses this limitation by preserving the PR-corrected anatomical signal while restoring texture statistics closer to GT. In simulated patients, both PRC-TP variants maintained recovery coefficients close to GT and substantially reduced the noise and CNR mismatch observed for PRC alone.

The texture validation supports the proposed function based signal dependent transfer mechanism. The input image showed the expected decrease in relative texture amplitude with increasing signal, and the fitted amplitude model captured this dependence. After MTRI, the restored texture reproduced the GT texture statistics as a function of signal and across regions with different PR induced signal changes. This agreement was statistical rather than voxelwise, which is the appropriate target because input and GT textures represent independent stochastic realizations. Radiomics analysis further showed that PRC-TP recovered texture sensitive image features attenuated in the smooth PRC prediction. The reduction in radiomics distance to GT, particularly for GLCM, GLRLM, and GLSZM features, suggests that texture restoration improves not only visual appearance but also the preservation of downstream image derived biomarkers. The similarity between PRC-TP$_0$ ($A(v) = 1$) and PRC-TP$_1$ indicates that the main recovery was driven by MTRI texture restoration, with limited additional effect from signal dependent amplitude modulation.
\begin{table*}[t]
\centering
\caption{Reference-free comparison of PRC-TP$_1$ between simulated and real clinical data. Changes are reported with respect to the original $^{82}$Rb input and with respect to PRC. Absolute changes are shown first, with relative changes in parentheses. Noise changes are expressed in percentage points (pp).}
\label{tab:prctp_sim_real_reference_free}
\renewcommand{\arraystretch}{1.15}
\resizebox{\textwidth}{!}{%
\begin{tabular}{lcccccc}
\toprule
\multirow{2}{*}{\textbf{}} 
& \multicolumn{3}{c}{\textbf{PRC-TP vs. $^{82}$Rb input}} 
& \multicolumn{3}{c}{\textbf{PRC-TP vs. PRC}} \\
\cmidrule(lr){2-4} \cmidrule(lr){5-7}
& \textbf{$\Delta$CR} 
& \textbf{$\Delta$CNR} 
& \textbf{$\Delta$Noise} 
& \textbf{$\Delta$CR} 
& \textbf{$\Delta$CNR} 
& \textbf{$\Delta$Noise} \\
\midrule

\textbf{Simulated} 
& $+0.21$ $(+10.99\%)$ 
& $-0.18$ $(-2.53\%)$ 
& $+1.79$ pp $(+19.74\%)$ 
& $+0.04$ $(+1.92\%)$ 
& $-5.88$ $(-45.87\%)$ 
& $+4.88$ pp $(+81.61\%)$ \\

\textbf{Real patient} 
& $+0.161$ $(+10.18\%)$ 
& $+1.679$ $(+31.70\%)$ 
& $-1.525$ pp $(-13.01\%)$ 
& $+0.017$ $(+0.98\%)$ 
& $-4.315$ $(-38.22\%)$ 
& $+4.095$ pp $(+67.13\%)$ \\

\bottomrule
\end{tabular}%
}
\end{table*}
Although GT data are not available for clinical acquisitions, the reference-free comparison between simulated and real $^{82}$Rb data showed consistent trends in Table~\ref{tab:prctp_sim_real_reference_free}. Relative to the original $^{82}$Rb images, PRC-TP increased contrast ratio by approximately 10\% in both simulated and clinical data (+10.99\% and +10.18\%, respectively). Similarly, when compared with PRC, PRC-TP
produced a substantial reduction in CNR together with a marked increase in image noise in both datasets. This behavior is expected, as texture re-injection restores stochastic fluctuations that are suppressed by the smooth PRC prediction. While this analysis does not constitute a quantitative validation, it provides supporting evidence that the trends observed under controlled simulation conditions are preserved in real $^{82}$Rb PET studies. Therefore, this clinical case should be interpreted as a preliminary feasibility and consistency assessment, rather than as clinical validation; evaluation in larger clinical cohorts will be required to assess robustness across patients, acquisition conditions, and reconstruction protocols.

Another consideration is that the quantitative validation relies on simulated GT images. Although the simulations were derived from real patient data and therefore preserve patient specific anatomical and activity distributions, the GT reference remains simulation based. In particular, cardiac and respiratory motion were not explicitly modeled in the present framework. Since motion may affect both apparent resolution and texture statistics in real $^{82}$Rb acquisitions, future work should evaluate PRC-TP under more realistic motion conditions and in larger clinical datasets.

An important characteristic of the proposed framework is that texture restoration is not based on synthetic texture generation. Unlike adversarial or generative approaches, which may introduce fluctuations that are not supported by the measured data, MTRI operates directly on the texture observed in the acquired PET image. As a result, the restored texture remains physically linked to the original acquisition and preserves its spatial correlations and statistical properties. The role of MTRI is therefore not to create new texture, but to adapt the amplitude of the measured texture to the signal level of the PR corrected prediction, reducing the risk of hallucinated image content while maintaining acquisition-consistent variability.

Because the central contribution of this work is the texture-preservation framework rather than the supervised PRC/N2N backbone itself, the ablation analysis focused on the MTRI components, including direct texture re-injection and signal-dependent modulation through $A(v)$. The training loss, sampling strategy, and network architecture were kept fixed across variants to isolate the effect of texture re-injection.

Overall, PRC-TP provides a practical strategy to combine PRC with acquisition-consistent texture preservation. By decoupling deterministic structure recovery from stochastic texture restoration, the framework improves apparent resolution while maintaining realistic PET texture, without relying on adversarial or generative models.

\section{Conclusion}
We proposed PRC-TP, a framework for PRC with explicit texture preservation for PET imaging. The method combines a tissue guided PRC network, an N2N derived smoothed input estimate, and a deterministic MTRI step that transfers acquisition fluctuations from the measured PET image to the PRC prediction. In simulated $^{82}$Rb cardiac PET, PRC-TP preserved the contrast recovery achieved by PRC while restoring noise, texture, and radiomics statistics closer to GT. The proposed function based MTRI formulation achieved near unity global texture amplitude agreement with GT without relying on explicit signal binning or generative texture synthesis. These results support PRC-TP as a practical framework for resolution recovery with texture preservation consistent with the original acquisition in high energy PET imaging.

\section{Acknowledgments}
This work was supported in part by the Spanish Ministry of Science, Innovation and Universities (MCIU) through the Agencia Estatal de Investigacion (AEI, 10.13039/501100011033) under Grants PID2021-126998OB-I00, TED2021-349130592B-I00, and PDC2022-133057-I00. It was further supported by the Comunidad de Madrid under the R\&D Technologies Program (Order 5696/2024) through the LUNABRAIN-CM Project (TEC-2024/TEC-43) and by National Institutes of Health (NIH), USA, under Grant R01EB033000. Robert J. Paneque-Yunta was supported by the Complutense University Predoctoral Fellowship (CT22/25), funded by the Complutense University of Madrid and Banco Santander. Cindy M. Solano-Cordero was supported by the Comunidad de Madrid through the Industrial Doctorate Program (IND2023/BMD-28275). The authors acknowledge the support of Siemens Medical Solutions USA, Inc. (Knoxville, TN, USA), and the provision of patient data by Washington University in St. Louis under IRB approval.

\bibliographystyle{IEEEtran}
\bibliography{ref}

\end{document}